\documentclass[%
 aip,
  pop, 
 amsmath,amssymb,
  reprint,%
]{revtex4-1}
\usepackage{CJK}

\usepackage{ulem}
\usepackage[T1]{fontenc}
\usepackage{calligra}

\usepackage{graphicx}
\usepackage{dcolumn}
\usepackage{bm}

\usepackage{times}
\usepackage{verbatim}
\usepackage{graphicx}
\usepackage{graphics,epsfig}

\usepackage{theorem}
\usepackage{makeidx}
\usepackage{amsmath}
\usepackage{epic}
\usepackage{amscd}

\usepackage{bbm}
\usepackage{xy}
\usepackage{amssymb}
\usepackage{epsfig}

\newcommand{\ignore}[1]{}  

\def \phi {\mbox{$\varphi$}}

\newsavebox{\astrutbox}
\sbox{\astrutbox}{\rule[-5pt]{0pt}{20pt}}

\newcommand{\bea}{\begin{eqnarray}}
\newcommand{\eea}{\end{eqnarray}}
\newcommand{\beq}{\begin{equation}}
\newcommand{\eeq}{\end{equation}}

\begin{document}
\begin{CJK*}{GB}{gbsn}

\preprint{AIP/123-QED}


\title[Spectral transfer in kinetic magnetized plasmas turbulence]{Fourier-Hankel/Bessel space absolute equilibria of 2D gyrokinetics}


\author{Jian-Zhou Zhu} 
\affiliation{\mbox{Department of Modern Physics, University of Science and Technology of China, Hefei, Anhui, 230026, China}}


\begin{abstract}
Two global invariants of two dimensional gyrokinetics are shown to be ``rugged'' (still conserved by the dynamics) concerning both Fourier and Hankel/Bessel Galerkin truncations. The truncations are made to keep only a finite range of wavenumber $\textbf{k}$ and the Hankel variable $b$ (or $z$ in the Bessel series). The absolute equilibria are used for the discussion of the spectral transfers in the configuration-velocity scale space of kinetic magnetized plasma turbulence. Some interesting aspects of recent numerical results, which were not well understood, are explained with more satisfaction.

%
\end{abstract}


\maketitle
\end{CJK*}

\thispagestyle{plain}

\section{\label{sec:Introduction}Introduction}

It has long been felt comfortable to sit in the chair of absolute equilibrium thinking about turbulence of neutral and conductive fluids \cite{Lee1952}. The basic idea is that the statistical solutions of the Galerkin truncated system with a subset of the Fourier modes may provide information about the turbulence physics such as the energy transfer (or cascade) dynamics and final states et al. In some cases the absolute equilibria spectra can even coincide with cascade ones \cite{LPPprl02,FPPS11}. Such an approach is still proving to be able to give new insights to Hall magnetohydrodynamic \cite{SMC08} and gyrokinetic \cite{ZhuHammett10} turbulence. The latter is the first in the kinetic framework with interesting new features which deserves some more introductive discussion here as the preparation of our work:

\subsection{the model}
For kinetic systems, such as the gyrokinetics, there are extra dimensions of velocities which should be dealt with appropriately. Zhu and Hammett \cite{ZhuHammett10} study the gyrokinetic equation for the electrostatic fluctuations in a slab geometry with uniform background magnetic field $\textbf{B}_0=B_0\textbf{z}$ reads (for a most updated historical review of the model even appropriate for non-plasma physicists, c.f., Krommes \cite{Krommes12})
\begin{eqnarray}\label{eq:gyronorm}
\frac{\partial g}{\partial t}
+ \left( \hat{\textbf{z}} \times \frac{\partial \langle \phi
  \rangle_{\textbf{R}} }{\partial \textbf{R}}
\right) \cdot \frac{\partial g}{\partial \textbf{R}}
=0
,
\end{eqnarray}
where $g(\textbf{R},v_{\perp})$ defined at the gyrocenter is a component of the fluctuating distribution $f$, $f=\exp\{-q\phi\}F_0+g+\langle \phi \rangle_{\textbf{R}}+h.o.t.$, deviating from the maxwellian $F_0$. Here $h.o.t.$ denotes the higher order terms in the gyrokinetic ordering and $\langle \cdot \rangle_{\textbf{R}}$ the gyroaverage around $\textbf{R}$: the average of any quantity $\Psi$ around a ring, of gyroradius $\rho$ surrounding the gyrocenter $\textbf{R}$,
perpendicular ($\perp$) to the magnetic field direction ($\parallel$), $\langle \Psi \rangle_{\textbf{R}}=\frac{ \int \Psi(\textbf{r}) \delta( \textbf{r}_{\parallel} - \textbf{R}_{\parallel} ) \delta[ |\textbf{r}_{\perp}-\textbf{R}_{\perp}|-\rho(\textbf{R}) ] d^{3}\textbf{r} } {\int \delta( \textbf{r}_{\parallel} - \textbf{R}_{\parallel} ) \delta[ |\textbf{r}_{\perp}-\textbf{R}_{\perp}|-\rho(\textbf{R}) ] d^{3}\textbf{r}}=\frac{ \int \Psi(\textbf{r}) \delta( \textbf{r}_{\parallel} - \textbf{R}_{\parallel} ) \delta[ |\textbf{r}_{\perp}-\textbf{R}_{\perp}|-\rho(\textbf{R}) ] d^{3}\textbf{r} } {2\pi \rho(\textbf{R})}$. [We explain that, throughout this article, if not particularly specified, $\int$ is used, for convenience, to denote actually the definite integral over the entire domain of the integration variable(s) - space or velocity. Using a Fourier representation $\Psi(\textbf{r}) = \sum_{\textbf{k}}
\exp(-i\textbf{k} \cdot \textbf{r}) \hat{\Psi}_\textbf{k}$, and considering a
straight magnetic field for simplicity here, we have $\langle \Psi
\rangle_{\textbf{R}}=\sum_\textbf{k} \exp(-i \textbf{k} \cdot
\textbf{R}) J_0(k_\perp \rho) \hat{\Psi}_{\textbf{k}}$, where $J_0$ is a Bessel
function.] The normalization by Plunk et al. \cite{GabeJFM10} brings conveniences for our discussion: The physical (dimensional) variables having subscript `p' is normalized as follows:
$t = t_{\mbox{\scriptsize{p}}}v_{\mbox{\scriptsize{th}}}/L$, $x = x_{\mbox{\scriptsize{p}}}/\rho_{th}$, $y = y_{\mbox{\scriptsize{p}}}/\rho_{th}$, $\phi = \phi_{\mbox{\scriptsize{p}}}\frac{q L}{T_0 \rho}$, $h = h_{\mbox{\scriptsize{p}}}\frac{v_{\mbox{\scriptsize{th}}}^3 L}{n_0 \rho}$ and $F_0 = F_{0\mbox{\scriptsize{p}}}v_{\mbox{\scriptsize{th}}}^3/n_0$.
The equilibrium density and temperature of the species of interest are $n_0$ and $T_0$; the thermal velocity is $v_{\mbox{\scriptsize{th}}} = \sqrt{T_0/m}$; the Larmor radius is $\rho_{th} = v_{\mbox{\scriptsize{th}}}/\Omega_{c}$ where the Larmor (cyclotron) frequency is $\Omega_{c} = qB/m$. This equation is closed by using the gyrokinetic quasi-neutrality
equation to determine the electrostatic potential
\begin{eqnarray}\label{eq:qn}
2\pi \int vdv \langle g \rangle_{\textbf{r}}=\alpha \varphi-\Gamma \varphi.
\end{eqnarray} 
For the plasma in a two dimensional (2D) cyclic box, with $v_{\parallel}$ being integrated out and $_{\perp}$ in $v_{\perp}$ omitted, the whole system in wavenumber space are
\begin{eqnarray}\label{eq:2DgkK}
\partial_t \hat{g}(\textbf{k},v)-\textbf{z}\times \sum_{\textbf{p}+\textbf{q}=\textbf{k}} \textbf{p}J_0(pv)\hat{\varphi}(\textbf{p})\cdot \textbf{q} \hat{g}(\textbf{q},v)=0
\end{eqnarray} and
\begin{eqnarray}\label{eqQNK}
\hat{\varphi}(\textbf{k})=\beta(\textbf{k}) \int vdv J_0(kv) \hat{g}(\textbf{k},v).
\end{eqnarray}
Here $\beta(\textbf{k})=\frac{2\pi}{\tau+1-\hat{\Gamma}(k)}$, and $\hat{\Gamma}(x)=I_0(x^2)e^{-x^2}$ is an exponentially-scaled
modified Bessel function. $I_0(x) = J_0(i x)$
and $\tau$ represents the shielding by the species which is treated as
having a Boltzmann response of some form [see, e.g., Plunk et al.\cite{GabeJFM10}; anisotropic response model, such as those respecting the zonal modes, can also be considered as in Zhu and Hammett\cite{ZhuHammett10}.]

\subsection{absolute equilibrium in $\textbf{k}-v$ space}
Fourier Galerkin truncation is defined by setting all Fourier modes beyond the wave number set $\mathbb{K}=\{\textbf{k}:k_{min}<|k|<k_{max}\}$ (, the summation over which will be denoted by $\tilde{\sum}$,) to be zero. Let us start with the 2D case. As Lee\cite{Lee1952} did for Euler equation for ideal fluid flow, we first observe that the dynamics of the ``gas'' composed by the real and imaginary parts of the Fourier modes, denoted by $\sigma$, satisfies the Liouville theorem. Actually ${\delta\dot{\sigma}\over \delta\sigma}=0$, where the dot represents the time derivative. Then, we proceed to find the canonical distribution. The only known rugged (still conserved after Fourier Galerkin truncation) invariants for the 2D gyrokinetic system are the (relative) entropy relevant quantity 
$$\mathrm{G}(v)=\int \frac{d^2\textbf{R}}{2V} g^2=\tilde{\sum}\frac{1}{2}|\hat{g}(\textbf{k},v)|^2$$ 
and the mean effective electrostatic potential ``energy'' 
$$\mathrm{E}=\int \frac{d^2\textbf{r}}{2V} [(1+\tau) \varphi^2-\varphi \Gamma \varphi]=\tilde{\sum}\frac{\pi}{2\beta(\textbf{k})} \left| \hat{\varphi}(\textbf{k}) \right|^2,$$ 
(see, e.g., Refs. \onlinecite{GabeJFM10,ZhuHammett10} and references therein) where $V$ is the volume (area) of the integration domain. Notice that $G(v)$ is a function of $v$, so we have now one plus a continuum of conserved quantities, which is the new feature of the problem, compared to earlier work for fluids.
Here the constant of motion $\mathcal{S}$ is formed by introducing
$\alpha_0$ and $\alpha(v)$, the ``(inverse) temperature parameters'' as the Lagrangian multipliers when  maximizing the Gibbs ensemble entropy:
\begin{eqnarray}\label{eq:H2Dc}
\mathcal{S} && =   \frac{1}{2} \tilde{\sum}_{\textbf{k}} \Bigg\{ \int \alpha(v) |\hat{g}(\textbf{k},v)|^2 dv+\alpha_0 2\pi  \beta(\textbf{k}) \cdot\nonumber\\
&& \cdot\int vdv J_0( kv) \hat{g}(\textbf{k},v)\int vdv J_0( kv) \hat{g}^{*}(\textbf{k},v) \Bigg\}
\end{eqnarray} with ``$^{\ast}$'' denoting the complex conjugate. 
 The canonical statistics of the Fourier Galerkin truncated system is then described by the distribution $Z^{-1} \exp\{-\mathcal{S}\},~Z=\int \!\! D\sigma  \exp\{-\mathcal{S}\}$.

The above is the calculation done by Zhu \cite{ZhuJFM11} which is the corresponding continuum limit (concerning velocity) of the results with discretized velocities by Zhu and Hammett \cite{ZhuHammett10}. We should remark that the electrostatic gyrokinetic equations constitute an integro-differential system. When the system are discretized, we should denote the corresponding solution with $\tilde{g}$ to distinguish it from the exact solution, $g$, of the original equations. In general, $\tilde{g}(v)\neq g(v)$. So, to be rigorous, for the discretized case in Ref. \onlinecite{ZhuHammett10} we should have replaced $\hat{g}(\textbf{k},v_i)$ with $\tilde{\hat{g}}(\textbf{k},v_i)$, as is the case of distinguishing the solution of the discretized equation (say, the difference equation) of a differential equation and the solution of the latter (people however often neglect such difference for it is not necessary for usual discussions.)

\subsection{the objective of this paper}
Although the calculation in the last subsection respects all possible rugged invariants of the original system, especially the results with discretized velocity correspond to the current gyrokinetic continuum codes, it is not convenient to discuss much the physics in velocity scale space (some finite discretization scale however still can be discussed as shown in Zhu \cite{ZhuJFM11} and has partly been verified numerically \cite{WatanabeZhu11}.) Especially, one could ask what would be the case if some other function spaces are used for the expansion of velocity (those in Zhu and Hammett \cite{ZhuHammett10} are the ``top hat'' bases as used in the continuum code.) This is the question we would address in this article.

In this paper, we will carry out the calculation of 2D gyrokinetic absolute equilibria in Fourier-Hankel/Bessel spaces and study the spectral issues based on the analytical results. We will show and compare several absolute equilibria of the gyrokinetic system with different treatments. All these absolute equilibria present some similar features for some aspects but also demonstrate different merits. We will then explain and comment on some recent numerical results.

\section{absolute equilibria in Fourier-Hankel/Bessel space}
As said, in the past \cite{ZhuHammett10,ZhuJFM11} we worked directly with velocity and respected all the local quadratic invariants $G(v)$, besides $E$. The results are precise and should be observed numerically. However, working directly with $v$, it is not convenient for studying the spectral transfers in velocity scale space, which requires introducing transformation of $v$ and using the appropriate global invariants.
 
\subsection{Using Fourier series and continuous Hankel transform}
It is interesting to note that the integration part in the right hand side of the quasi-neutrality condition, Eq. (\ref{eqQNK}), is exactly the definition of the Hankel transform which will be denoted with a breve, $\breve{}$. That is, 
\begin{eqnarray}
\hat{\varphi}(\textbf{k})=\beta(\textbf{k}) \int vdv J_0(kv) \hat{g}(\textbf{k},v)=\beta(\textbf{k})\breve{\hat{g}}(\textbf{k},k),
\end{eqnarray}
which suggests the application of Hankel transform for $v$. So, further doing Hankel transform for the second argument $v$ (to $b$,) we have
\begin{eqnarray}\label{eq:gkb}
 \frac{\partial \breve{\hat{g}}(\textbf{k},b)}{\partial t}&=&\int vdv J_0(bv) \textbf{z}\times \sum_{\textbf{p}+\textbf{q}=\textbf{k}} \!\!\!\!\! \textbf{p}J_0(pv)\hat{\varphi}(\textbf{p})\cdot \\\nonumber
&&\cdot \textbf{q} \int wdw J_0(vw)\breve{\hat{g}}(\textbf{q},w).
\end{eqnarray}

As $G(v)$ is conserved for all $v$, it is equivalent to say that, for any reasonable (test) function $T(v)$, $W_T=\int T(v) G(v)vdv$ is conserved. In the scale, $k-b$, space, the invariants can be represented as 
\begin{eqnarray}\label{eq:E}
E=\frac{\pi}{2}\sum_{\textbf{k}}\beta(\textbf{k})\int \delta(b-k)|\breve{\hat{g}}(\textbf{k},b)|^2 db
\end{eqnarray}
and 
\begin{eqnarray}\label{eq:WT1}
W_T &=& \int [\sum_{\textbf{k}} \int \breve{\hat{g}}(\textbf{k},b)J_0(bv)bdb \nonumber \\
 && \int \breve{\hat{g}}^{\star}(\textbf{k},b')J_0(b'v)b'db']T(v)vdv.
\end{eqnarray} 
We observe that $E$ is rugged with both Fourier- and Hankel-Galerkin truncation (keeping only a subset of $\textbf{k}$ and $b$), and this is the same for $W$ when $T=1$. The question is whether we would miss any other rugged invariants with the extra Galerkin truncation for b. We believe the answer is ``No," which can be shown following Kraichnan \cite{KraichnanJFM73}: Integrating out $v$ first in Eq. (\ref{eq:WT1}) we have
\begin{eqnarray}\label{eq:WT2}
W_T = [\sum_{\textbf{k}} \int \breve{\hat{g}}(\textbf{k},b)bdb  \int \breve{\hat{g}}^{\star}(\textbf{k},b')b'db']J(b,b'),
\end{eqnarray}
where $J(b,b')=\int J_0(bv)J_0(b'v)T(v)vdv$.
Now, introduce the Galerkin truncation (for however \textit{continuous} $b$) formally in the same way as for discrete $\textbf{k}$, that is, forcing $\breve{\hat{g}}(\textbf{k},b)=0$ for $b\not\in\mathbb{B}=\{b: b_{min}<b<b_{max}\}$. Then for general $J(b,b')$, as $\frac{\partial \breve{\hat{g}}(\textbf{k},b)}{\partial t}$ is generally not zero even for $b\not\in\mathbb{B}$, $\frac{d \tilde{W}_T}{d t}$ is not zero. Here,
\begin{eqnarray}\label{eq:WT3}
\tilde{W}_T = [\tilde{\sum}_{\textbf{k}} \tilde{\int} \breve{\hat{g}}(\textbf{k},b)bdb  \tilde{\int} \breve{\hat{g}}^{\star}(\textbf{k},b')b'db']J(b,b')
\end{eqnarray} 
as is the meaning of Fourier-Hankel Galerkin truncation. The reason is that for $\breve{\hat{g}}$ supported only by $\mathbb{B}$, $\frac{d W_T}{d t}$ ($=0$) contains not only $\frac{d \tilde{W}_T}{d t}$ but also the integration over $b$ and $b'$ of $\partial_t \breve{\hat{g}}(\textbf{k},b)b \breve{\hat{g}}^{\star}(\textbf{k},b')b'J(b,b')$, which is not zero in general, with $b'$ in $\mathbb{B}$ but $b$ not. In such case, $\frac{d \tilde{W}_T}{d t}\equiv0$ only when $J(b,b')\equiv0$ for one and only one of $b$ and $b'$ is in $\mathbb{B}$: As $b_{min}$ and $b_{max}$ are quite arbitrary, the only nontrivial generic function satisfying such a property is the Dirac function, that is $J(b,b')=\delta(b-b')/b$ which corresponds to $T(v)=1$. The trivial one is $J(b,b')\equiv0$ uniformly over the full domain, or equivalently $T(v)\equiv0$ which is physically not interesting. It is worthy to point out that when $T(v)=\delta(u-v)/v$, $J(b,b')=J_0(bu)J_0(b'u)$ and that $W_T=G(u)$, but for the reason just explained $G(u)$ is not a rugged invariant. Actually, for given $b\in\mathbb{B}$, the only solution of $J(b,b')=0$ for all $b'$ not in $\mathbb{B}$ seems to be $T(v)=0$ or $T(v)=1$. So, we believe $E$ and $$W=\sum_{\textbf{k}} \int |\breve{\hat{g}}(\textbf{k},b)|^2bdb$$ are the only rugged invariants with the Galerkin truncation of both $\textbf{k}$ and $b$.

With the arguments given in the above, we conclude the canonical distribution be 
\begin{eqnarray}
\sim && \exp\Big{\{}\big{[}-\gamma_E\tilde{\sum}_{\textbf{k}}\beta(\textbf{k})\tilde{\int} \delta(b-k)|\breve{\hat{g}}(\textbf{k},b)|^2 db -\nonumber \\
&&-\gamma_W \tilde{\sum}_{\textbf{k}} \tilde{\int} |\breve{\hat{g}}(\textbf{k},b)|^2bdb\big{]}/2\Big{\}},
\end{eqnarray}
which leads to $\langle|\check{\hat{g}}(\textbf{k},b)|^2\rangle=\frac{1}{\gamma_W b+\gamma_E \beta(\textbf{k})\delta_{k,b}}$ with $\delta_{k,b}$ acquiring 1 for $k=b$, and 0 otherwise. The calculation with discretization of $b$, say, that parameterized by uniform lattice size $\Delta b$, gives $\langle|\check{\hat{g}}(\textbf{k},b_i)|^2\rangle=\frac{1}{\Gamma_W b_i \Delta b+\Gamma_E \beta(\textbf{k})\delta_{i_k,i}}$ ($k$ falls into the lattice of $b$ indexed by $i_k$) with $\Gamma_W \Delta b \to \gamma_W$ and $\Gamma_E \to \gamma_E$, and, $\delta_{i_k,i} \to \delta_{k,b}$. The spectra densities are then
\begin{eqnarray}
E(\textbf{k},b) &=& \frac{\pi}{2}\beta(\textbf{k}) \delta(b-k)\langle |\breve{\hat{g}}(\textbf{k},b)|^2 \rangle  \nonumber \\
&=& \frac{\pi}{2} \frac{\beta(\textbf{k})\delta(b-k)}{\gamma_W b+\gamma_E \beta(\textbf{k})\delta_{k,b}}\label{eq:Ekb}\\
  W(\textbf{k},b) &=& b\langle |\breve{\hat{g}}(\textbf{k},b)|^2 \rangle  =  \frac{b}{\gamma_W b+\gamma_E \beta(\textbf{k})\delta_{k,b}}\label{eq:Wkb}
\end{eqnarray}
The mean energy is then
\begin{eqnarray}\label{eq:meanE}
\langle \tilde{E}\rangle &=&\tilde{\sum}_{\textbf{k}}\tilde{\int} db E(\textbf{k},b) = \tilde{\sum}_{\textbf{k}} \mathcal{E}(\textbf{k},k),
\end{eqnarray}
with
\begin{equation}\label{eq:Ekk}
\mathcal{E}(\textbf{k},k)=\frac{\pi}{2}\frac{\beta(\textbf{k})}{\gamma_W k+\gamma_E \beta(\textbf{k})}
\end{equation}
for $k\leq k_{max}$ and $k\leq b_{max}$; $\mathcal{E}(\textbf{k},k)$ is zero for $k> b_{max}$ or $k> k_{max}$ and is $\mathcal{E}(\textbf{k},b_{max}^-)$ - the right limit - for $k=b_{max}$ and $k\leq k_{max}$, although the value of $\int_{-\infty}^{b_{max}}\delta(b-b_{max})db$ is undefined (one however could try to define it through some particular nascent delta function.)

\subsection{Using Fourier and Bessel series}
As in computer simulations and in finite experiments, the velocity is finite, it may also be useful to transform velocity variable $v$ to Hankel space with Bessel series by taking $v$ to be bounded. Suppose velocity is bounded by $V$, we then have 
\begin{eqnarray}
\hat{g}(\textbf{k},v)&=&\sum_{z} 2V^{-2}[J_1(z)]^{-2}\breve{\hat{g}}(\textbf{k},z) J_0(z v/V)\nonumber \\
&=&\sum_{z} J_z(v)\breve{\hat{g}}(\textbf{k},z),
\end{eqnarray}
with $z$ being the zeros of $J_0$; and, from the orthogonality relationship
\begin{eqnarray}
\int_0^V J_\alpha(\frac{x z_m}{V})\,J_\alpha(\frac{x z_n}{V})\,x\,dx = V^2 \frac{\delta_{mn}}{2} [J_{\alpha+1}(z_n)]^2,
\end{eqnarray}
we have
\begin{eqnarray}\label{eq:disgkb}
&& \frac{\partial \breve{\hat{g}}(\textbf{k},z_n)}{\partial t}=\int_0^{V}vdv J_0[\frac{z_n v}{V}] \textbf{z}  \times \sum_{\textbf{p}+\textbf{q}=\textbf{k}} \!\!\!\!\! \textbf{p}J_0(pv) \varphi(\textbf{p}) \cdot \nonumber \\
&&\cdot \textbf{q} \sum_{m} J_{z_m}(v)\breve{\hat{g}}(\textbf{q},z_m)
\end{eqnarray}
and the quasi-neutrality condition
\begin{eqnarray}
\!\!\!\!\!\! \hat{\varphi}(\textbf{k})&&=\beta(\textbf{k}) \!\! \int \!\! vdv J_0(kv) \sum_{z} \frac{2}{V^{2}}\frac{1}{J_1^2(z)}\breve{\hat{g}}(\textbf{k},z) J_0(\frac{zv}{V})\nonumber \\
&&=\sum_{z} B(\textbf{k},z)\breve{\hat{g}}(\textbf{k},z)
\end{eqnarray}

With the similar arguments as in the continuous Hankel transform case, we can see that now only $W$ is rugged with regard to both Fourier and Bessel Galerkin truncation and $$\tilde{W}=\tilde{\sum}_{\textbf{k}}\tilde{\sum}_{z} 2V^{-2}J_1^{-2}(z)|\breve{\hat{g}}(\textbf{k},z)|^2.$$ The ruggedness of $E$ is lost due to the fact that $B(\textbf{k},z)$ is in general distributed (over z). The ruggedness of $E$ would be recovered when $B(\textbf{k},z)=\beta(\textbf{k})\delta_{z,z_{(k)}}$ and $\hat{\varphi}(\textbf{k})=\beta(\textbf{k})\breve{\hat{g}}(\textbf{k},z_{(k)})$ where $z_{(k)}$ is some zero of $J_0$: This happens when $k=\frac{z_{(k)}}{V(\textbf{k})}$. That is, the velocity bound $V$ is taken to be $\textbf{k}$ dependent, which was initiated courageously by Plunk and Tatsuno \cite{GabeTomoPRL11}. In this way, 
\begin{eqnarray*}
\tilde{E}=\tilde{\sum}_{\textbf{k}} \frac{\pi}{2} \beta(\textbf{k})|\breve{\hat{g}}(\textbf{k},z_{(k)})|^2=\tilde{\sum}_{\textbf{k},z} \frac{\pi}{2} \beta(\textbf{k})|\breve{\hat{g}}(\textbf{k},z)|^2\delta_{z,z_{(k)}}.
\end{eqnarray*} 
Note that the dynamical equation (\ref{eq:disgkb}) needs to be changed accordingly.

Before going further, we digress to comment that the introduction of wavenumber dependent truncation of velocity amplitude $V(\textbf{k})$ brings convenience as well as subtleties. The obvious point is that $V(\textbf{k})$ can not be normalized uniformly, as inappropriately done by Plunk and Tatsuno \cite{GabeTomoPRL11} in their Eq. (5), and that it will affect the relation between the spectra of $E$ and $W$ [the relation given in Plunk and Tatsuno's \cite{GabeTomoPRL11} Eq. (6) is then not correct - see below] which is critical for the Fjortoft argument tried in Ref. \onlinecite{GabeTomoPRL11}. Physically, in natural phenomena, laboratory experiment or numerical simulations, the upper bound of velocity is unknown or uncontrolled, so it is not very clear what exactly the relevance of the $\textbf{k}$ dependent upper bound of velocity besides the mathematical convenience (though it is obvious that velocity fluctuations at various spacial scales are different.) Nevertheless, we can try to continue the absolute equilibrium calculation to make things clearer.

The absolute equilbrium distribution for such wavenumber dependent upper bound of velocity amplitude
is $\sim \exp\{-(\alpha_E \tilde{E} + \alpha_W \tilde{W})/2\}$
which gives the spectral density of $\tilde{E}$ and $\tilde{W}$:
\begin{eqnarray}\label{eq:Ekz}
E(\textbf{k},z) & \triangleq\langle & \frac{\pi}{2} \beta(\textbf{k}) \delta_{z,z_{(k)}}|\breve{\hat{g}}(\textbf{k},z)|^2 \rangle \nonumber \\
 &=& \frac{\pi \delta_{z,z_{(k)}}\beta(\textbf{k})}{\alpha_E \pi \beta(\textbf{k})\delta_{z,z_{(k)}}+ 4 \alpha_W V^{-2}(\textbf{k}) J_1^{-2}(z)}
\end{eqnarray}
with $z_{(k)}=kV(\textbf{k})$ and
\begin{eqnarray}\label{eq:Wkz}
&&  W(\textbf{k},z)\triangleq\langle 2V^{-2}(\textbf{k})J_1^{-2}(z)|\breve{\hat{g}}(\textbf{k},z)|^2 \rangle \\\nonumber
&&=\frac{4}{\pi\alpha_E\beta(\textbf{k})J_1^2(z)\delta_{z,z_{(k)}}V^{2}(\textbf{k})+4\alpha_W}
\end{eqnarray}
with the similar considerations below Eq. (\ref{eq:Ekk}). 
By definition
\begin{equation}
\begin{pmatrix}
  \langle E \rangle \\
  \langle W \rangle \\
\end{pmatrix}
=\tilde{\sum}_{\textbf{k}}\tilde{\sum}_{z}
\begin{pmatrix}
  E(\textbf{k},z) \\
  W(\textbf{k},z) \\
\end{pmatrix}
\end{equation}
 
\subsection{on the spectral transfer}

Note that for the finite velocity and discrete Bessel series case
\begin{eqnarray}\label{eq:EWkz}
W(\textbf{k},z_{(k)})=\frac{4}{\pi} V^{-2}(\textbf{k})J_1^{-2}(z_{(k)})E(\textbf{k})/\beta(\textbf{k}),
\end{eqnarray} 
with $E(\textbf{k})=E(\textbf{k},z_{(k)})$.
And, from Eqs. (\ref{eq:meanE}), (\ref{eq:Ekb}) and (\ref{eq:Wkb}) for the continuous Hankel transform case, we have 
\begin{eqnarray}\label{eq:EWkb}
W(\textbf{k},k)=\frac{2k}{\pi\beta(\textbf{k})}\mathcal{E}(\textbf{k},k).
\end{eqnarray} 
Note that these relations are valid in general, not only for the absolute equilibrium spectra.
The Fjortoft arguments concerning the constraints of spectral transfers can be carried over, {\it mutatis mutandis}, as pioneered by Plunk and Tatsuno \cite{GabeTomoPRL11}, but with flawed analysis and inappropriate statements as commented by the author \cite{CZ}: For example, the $\textbf{k}$ dependence of the upper bound $V(\textbf{k})$ could affect the relations and should have been taken account [even though they were discussing the large $k$ limit where $\beta(\textbf{k})$ was approximated as constant and $J_1^{-2}(z_{(k)})$ was approximated as $z_{(k)}$,] and the constraints can not tell the directions of the transfers, among others.

The balance equation for studying the energy spectral transfer can be derived from Eqs. (\ref{eq:gkb}) and (\ref{eq:meanE}).
\begin{eqnarray}\label{eq:balance}
\frac{d\mathcal{E}(\textbf{k},k)}{dt}&=& 2\beta^2(\textbf{k})\sum_{\textbf{p}+\textbf{q}=\textbf{k}}\int wdw \frac{1}{2\pi\Delta(k,p,w)}\nonumber \\ 
&&\langle \breve{\hat{g}}(\textbf{p},p)\breve{\hat{g}}(\textbf{q},w)\breve{\hat{g}}^*(\textbf{k},k) \rangle \textbf{z} \times \textbf{p} \bullet \textbf{q},
\end{eqnarray}
where $$\Delta(k,p,w)=1/[2\pi \int vdv J_0(kv)J_0(pv)J_0(wv)]$$ is the area of the triangle formed with legs of lengths $k$, $p$ and $w$: the area is taken to be infinite if the three legs are not appropriate to close a triangle \cite{GervoisNaveletJMP84}. The right hand side of Eq. (\ref{eq:balance}) is called the transfer rate function $T(\textbf{k})$ satisfying $\sum_{\textbf{k}}T(\textbf{k})=0$. One should calculate the flux from the transfer rate function to determine the energy flow in the $\textbf{k}-b$ space and in general it is a very difficult mission (but Kraichnan \cite{Kraichnan2D67} was able to estimate the signs of fluxes for some particular cases of 2D Navier-Stokes turbulence.) Similarly is the case for $W$ transfer analysis, and also for the velocity bounded case. In this paper, we don't attempt to estimate the value of sign of the transfer rate or flux. Writing the balance equation down, we just mean to remark that there are details in the transfer which are not all well described by ``macroscopic'' arguments, such as the Fjortoft constraints or the tendency of relaxation to absolute equilibria. For example, Eq. (\ref{eq:balance}) shows that the energy transfer is accomplished by the interactions of the triangles relating both the configuration- and velocity-scale spaces with diagonal ($\textbf{p}-p$) modes, which should be respected by any physical conjectures.

The absolute equilibrium spectra [Eq. (\ref{eq:Ekk}) and similarly Eq. (\ref{eq:Ekz})] is useful for predicting the transfer directions and large-scale structure formulation. This is very similar to the case of 2D Naver-Stokes turbulence, except that we now have the extra dimension of velocity fluctuation scales. Just as Kraichnan discussed for 2D fluid turbulence \cite{Kraichnan2D67}, the negative temperature [$\gamma_E$ in Eq. (\ref{eq:Ekk}) and $\alpha_E$ in Eq. (\ref{eq:Ekz})] state shows the energy will peak at the lowest modes [note that $\beta(\textbf{k})$ is a decrease function of $k$]. The relevance of equilibrium statistical mechanics with negative temperature to large scales of 2D turbulence has various supports starting from Onsager \cite{Onsager49} and recently from the Miller-Robert theory \cite{Robert}, and also from the perturbation from $\frac{4}{3}$D absolute equilibrium by L'vov, Pomyalov and Procaccia \cite{LPPprl02}. As Kraichnan discussed \cite{Kraichnan2D67}, the negative temperature may due to the large ratio of $E$ over $W$: In (plasma) turbulence, the external forcing could inject large amount of energy into the system and/or the collisional dissipation of $W$ stronger than that of $E$ could lead to such large $E/W$.

The possibility of anisotropic $\beta(\textbf{k})$, as discussed in Zhu and Hammett \cite{ZhuHammett10} with the consideration of zonal flows, will not only change the final absolute equilibrium spectra but also the relations between them and that the transfer dynamics in scale space. Note that the $\beta(\textbf{k})$, and then its anisotropy, enters in the absolute equilibrium spectra, even for $W$, only when $k=p$.

\section{discussion} 
The main resulted formulae for the discussion of spectral transfers are the absolute equilibrium spectral densities Eqs. (\ref{eq:Ekb},\ref{eq:Wkb}) and (\ref{eq:Ekz},\ref{eq:Wkz}). The former ones are even simpler and cleaner than the latter which are ``contaminated'', with $V(\textbf{k})$ and $J_1(z)$, by the effects of scale dependent truncation of velocity amplitude. In the large $z$ limit and with $V$ almost uniform (Plunk and Tatsuno \cite{GabeTomoPRL11} actually took such limits without justification, especially for the uniform $V$) the two cases are similar, but they could also be drastically different beyond these limits. Usually, the theory by taking the upper bound to be infinity may be sufficient and convenient, as in general the realistic largest velocity is large enough.

We have seen that different treatments of the system may lead to different versions of absolute equilibria. None of them is the exact turbulence solution, however close it could be. Which one is physical or physically relevant, or how the physical relevance could be made is the critical issue. Concerning the ``as simple as possible, but not simpler'' quote (generally attributed to Einstein,) a remark about the theory follows. Though all absolute equilibria may be precise regarding the respective expansion and truncation, for studying the spectral transfers in both configuration and velocity scale space the one with the continuous Hankel transform of velocity seems to have been ``as simple as possible'' and the one with Fourier-Bessel series using $\textbf{k}$ dependent upper bound of velocity is ``simple'', while the one with Fourier-Bessel series using a uniform upper bound of velocity is ``simpler''; the one without further transformation of velocity in Zhu and Hammett \cite{ZhuHammett10} however is not simple enough for this purpose. Of course, different versions of absolute equilibria have their own merits and comparisons among them may bring more insights. For example, the ones working directly with velocity, with or without discretization of velocity \cite{ZhuJFM11}, can help to quantify the numerical discretization (the current continuum codes use exactly the same discretizations as in the analysis) effects and its possible physical consequences; and, the ones with or without upper bound of velocities as presented here can help to quantify the effects of finiteness of velocities. Actually, turbulence is so complicated, so we need various exact (statistical) solutions of relevant systems (derived from the original system) to gain more insights into it. For example, it may be a good strategy to start from the theory respecting all invariants of the collisionless (inviscid) system \cite{Robert} for a better knowledge of the possible physical relevance of the dropped ones.

Different versions of absolute equilibria have their common features, such as the negative temperature states with energy condensation at lowest modes (which should not be very surprising as the 2D gyrokinetics plasma should reach the hydrodynamic limit, where such features have been well documented, in the cold ion limit,) but also differences. How the differences of the spectra behavior would be physically relevant to predict particular features of real turbulence is subtle and to be examined. For example, the author \cite{ZhuJFM11} discussed the effect of finite discretization might lead to also condensation of $G(v)$ at lowest modes, which seems to have been confirmed \cite{WatanabeZhu11} by numerical simulation of the absolute equilibrium ensemble but however still needs to be further examined in real (``real'' only numerically with collision operators and possibly also forcing) turbulence. It is unclear what the physical relevance (if any) is when the upper bound of velocity truncation is brutally taken to be uniform so that the ruggedness of $E$ is lost (the absolute equilibria simply correspond to the case with $\alpha_E=0$.)

In three dimension (3D) dynamics, parallel electric field acceleration, parallel advection et al., besides the slow gyroaveraged $\textbf{E} \times \textbf{B}$ drift, take effect. The physical mechanisms of spectral transfer are a lot more. In the standard gyrokinetic ordering, other parallel nonlinearity are of higher order and the parallel dynamics adds only linear terms. While the unchanged nonlinearity preserves $E$ and $W_T$ separately, the parallel motions enter to combine them into a single invariant \cite{ZhuHammett10}. Extra techniques are necessary to extract the velocity scales of the mixed linear and nonlinear phase mixing for investigating into the spectral transfer, though insights of transfers in configuration space scales are relatively easier \cite{ZhuHammett10}. It is noted earlier \cite{ZhuJFM11} that the 3D absolute equilibrium seems to be more robust concerning the finite velocity discretization (quantization), we wonder whether this could be the case when the quantization is introduced with finite amplitude of the upper bound $V$.

Finally, the Fjortoft argument as generalized by Plunk and Tatsuno \cite{GabeTomoPRL11} constraints the possible transfer directions of one of the variables when the other one's direction is already known, but it does not tell where the system should start and continue going. The absolute equilibria, containing already the information of the constraints from the Fjortoft argument, tells the system the arrow of time. So, the absolute equilibria analysis may be ready to explain many aspects of the spectral transfer results of the current numerical simulations, such as those in Ref. \onlinecite{GabeTomoPRL11}. For example, the propagation directions of the modes are explained with the tendency of relaxation to the absolute equilibria, especially before the regime when and where collision and/or forcing (if exists) operator takes effect. And, especially, our Eq. (\ref{eq:Wkb}) shows that $W(\textbf{k},b)$ is symmetric about the line $k=b$ and then $\sum_{|\textbf{k}|=k}W(\textbf{k},b)$ should have accumulate more on the $k$ axis side where more $\textbf{k}$ modes sit, which should persist in the nonequilibrium spectrum without particular non-symmetric stirring: This simple result seems to explain the non-symmetric behavior in Fig. 6 of Tatsuno et al. \cite{TatsunoJPFRS10} where larger values are on the $k$ axis side (the same phenomena has been found by Watanabe. \cite{WatanabeZhu11}) We expect that, if $\sum_{|\textbf{k}|=k}W(\textbf{k},b)/N(k)$, with $N(k)$ being the number of modes on the $k$ shell, is plotted, the figure should be symmetric about the line $k=b$: This is not easily seen from the transfer rate function in the corresponding balance equation, but we can imagine that, as for any symmetric truncations of $k$ and $b$ the corresponding absolute equilibria is symmetric, the transfer should also be symmetric.

\begin{acknowledgments}
This work is partly supported by ``the Fundamental Research Funds for the Central Universities'': WK 2030040016.
The author thanks many colleagues for discussing relevant things during the course of this work, especially G. Plunk, P. Diamond, D. Escande, U. Frisch, G. Hammett, T. Tatsuno, R. Waltz, T.-H. Watanabe.
\end{acknowledgments}


\begin{thebibliography}{10}%
\bibitem{Lee1952}
T.-D. Lee, Q. Appl. Math. {\bf 10}
, 69 (1952).

\bibitem{LPPprl02}
V. S. L'vov, A. Pomyalov, I. Procaccia, Phys. Rev. Lett. {\bf 89}, 064501 (2002).

\bibitem{FPPS11}
U. Frisch et al., arXiv:1108.1295v1 [nlin.CD]

\bibitem{SMC08}

Sergio Servidio, William H. Matthaeus, and Vincenzo Carbone, Phys. Plasmas {\bf 15}, 042314 (2008)

\bibitem{ZhuHammett10}
J.-Z. Zhu and G. W. Hammett, Phys. Plasmas {\bf} 17, 122307 (2010)

\bibitem{Krommes12}
J. A. Krommes, Annual Review of Fluid Mechanics, {\bf 44}, to be published (2012).

\bibitem{GabeJFM10}
G. Plunk et al., J. Fluid Mech., {\bf 664}, 407 (2010).

\bibitem{ZhuJFM11}
J.-Z. Zhu, arXiv:1008.0330v3 [nlin.CD].

\bibitem{WatanabeZhu11}
T.-H.Watanabe and J.-Z. Zhu, Private communication (2011).

\bibitem{KraichnanJFM73}
R. H. Kraichnan, J. Fluid Mech., {\bf 59}, 745 (1973). 

\bibitem{GabeTomoPRL11}
G. Plunk and T. Tatsuno, Phys. Rev. Letters, {\bf 106}, 165003 (2011)

\bibitem{CZ}
J.-Z. Zhu, arXiv:1105.1593v4 [nlin.CD].

\bibitem{GervoisNaveletJMP84}
A. Gervois and H. Navelet, J. Math. Phys. {\bf 25}, 3350 (1984).

\bibitem{Kraichnan2D67}
R. H. Kraichnan, Phys. Fluids, {\bf 10}, 1417 (1967).

\bibitem{Onsager49}
L. Onsager, Nuovo Cimento, {\bf 6}, 279 (1949).

\bibitem{Robert}
R. Robert "Statistical Hydrodynamics (Onsager Revisited)", in: Handbook
of mathematical fluid dynamics, Volume 2, Eds. Susan Friedlander, Denis Serre (Gulf Professional Publishing, 2003)

\bibitem{TatsunoJPFRS10}
T. Tatsuno et al., J. Plasma Fusion Res. SERIES {\bf 9}, 509 (2010)


\end{thebibliography}
\end{document}